\documentclass[11pt,a4paper]{article}
\usepackage{amsmath,amssymb,titling,authblk}
\usepackage{amsthm}
\usepackage[top=2.3cm,right=2.3cm,left=2.3cm,bottom=2.3cm]{geometry}
\usepackage{graphicx}
\usepackage{color} 
\usepackage{slashed}
\usepackage[pdfstartview=FitH,colorlinks=true,linkcolor=blue,anchorcolor=red,citecolor=magenta,urlcolor=blue]{hyperref}
\usepackage{amscd}
\usepackage[normalem]{ulem}
\usepackage{appendix}
\usepackage{bbold}
\usepackage{mathrsfs}
\usepackage{pdfsync}
\usepackage{bbm}
\usepackage{bm}
\usepackage[arrow,matrix,curve]{xy}
\usepackage{bbding}
\usepackage{wasysym}
\usepackage{booktabs}
\usepackage{siunitx}
\usepackage{cite}
\usepackage{epsf}
\usepackage{epsfig}
\usepackage{wrapfig}
\usepackage[utf8]{inputenc}
\usepackage{subcaption}

\DeclareCaptionFormat{custom}
{%
    \textbf{#1#2}\textit{\small #3}
}
\allowdisplaybreaks[4]
\numberwithin{equation}{section}

\definecolor{verde}{cmyk}{.83,.21,1,.08}
\definecolor{darkorchid}{rgb}{0.6, 0.2, 0.8}
\definecolor{darkgreen}{rgb}{0,.5,0}

\def\({\left(}
\def\){\right)}
\def\[{\left[}
\def\]{\right]}

\newcommand{\ii}{\mathrm{i}}

\newcommand{\dd}{\mathrm{d}}

\newcommand{\be}{\begin{equation}}
\newcommand{\ee}{\end{equation}}
\newcommand{\bea}{\begin{eqnarray}}
\newcommand{\eea}{\end{eqnarray}}

\newcommand{\la}{\label}

\begin{document}

\title{Symmetries and Conservation Laws in Lie-Poisson Electrodynamics
}

\author[1,2]{M. A. Kurkov}

\affil[ ]{}

\affil[1]{\textit{\footnotesize Dipartimento di Fisica ``E. Pancini'', Universit\`a di Napoli Federico II, Complesso Universitario di Monte S. Angelo Edificio 6, via Cintia, 80126 Napoli, Italy.}}
\affil[2]{\textit{\footnotesize INFN-Sezione di Napoli, Complesso Universitario di Monte S. Angelo Edificio 6, via Cintia, 80126 Napoli, Italy.}}
\affil[ ]{}
\affil[ ]{\footnotesize e-mail: \texttt{max.kurkov@gmail.com}}
\maketitle
\begin{abstract}\noindent
Lie-Poisson electrodynamics (LPE) is a non-Abelian and nonlinear deformation of usual electrodynamics, where the gauge algebra is defined through a Lie-algebra-type Poisson bracket on space-time. This field-theoretical model arises from noncommutative U(1) gauge theory in the semiclassical limit. We focus on the geometric approach to LPE in the absence of charged matter. 
We establish a non-trivial field redefinition which, under mild technical assumptions, maps the LPE dynamics to that of Maxwell theory.   {We emphasise that despite the dynamical equivalence, these two theories are not physically equivalent, since the corresponding local gauge-invariant observables are different. We show that our field redefinition is actually a Seiberg-Witten map between LPE and Maxwell theory.}
Using this map, for any symmetry of the Maxwell action, we construct generators of LPE symmetries and the corresponding conserved currents. In particular, we obtain deformed Poincaré transformations, thereby solving the problem of constructing space-time symmetries in LPE.   The deformed  {symmetries} are constructed as active transformations of the gauge field. In the outlook, we propose a natural quantization prescription for LPE based on our field redefinition.
\end{abstract}

\section{Introduction}
Gauge theories are the cornerstone of elementary particle physics.  For many decades, the canonical example has been Yang-Mills theory, characterized by the  algebra of infinitesimal gauge transformations
\be
[\delta^{\mathrm{YM}}_{{\bf f}},\delta^{\mathrm{YM}}_{{\bf g}}] = \delta^\mathrm{YM}_{[\bf f,\bf g]}
\ee
with $\bf f$ and $\bf g$ being Lie-algebra-valued infinitesimal parameters. Although the strong and weak interactions are essentially non-Abelian, the electromagnetic sector is described by an Abelian gauge theory, whose predictions have been confirmed with remarkable precision. One may wonder, however, whether this Abelian nature is exact, or whether it could be violated at extremely short distances inaccessible to present experiments, for instance, at the Planck scale.

A natural candidate for deformed electrodynamics is Poisson gauge theory, also known as Poisson electrodynamics~\cite{Kupriyanov:2023qot}. It is a non-Abelian deformation of $U(1)$ gauge theory where the infinitesimal transformations obey
\be
[\delta_f,\delta_g]=\delta_{\{f,g\}} ,
\ee
and $\{f,g\}$ is a given Poisson bracket on space-time. 
On the one hand, this formalism was born from noncommutative geometry (NCG), as a semiclassical limit of noncommutative $U(1)$ gauge theory, see~\cite{Kupriyanov:2020sgx,Kupriyanov:2023gjj} for details. For a general review on noncommutative gauge theories, we refer to~\cite{Hersent:2022gry}. 
On the other hand, Poisson electrodynamics is self-contained and can be considered per se.

A well-developed class of Poisson gauge models is given by Lie-Poisson electrodynamics (LPE), where the space-time coordinates close a Lie algebra under the Poisson bracket~\cite{Kupriyanov:2023gjj}:
\be
\{x^\mu,x^\nu\}=C^{\mu\nu}_\lambda x^\lambda {.}  \la{xxcomm}
\ee
 {Generally} speaking, the indices can be either spatial or temporal.
Various noncommutativities that have attracted significant attention in the literature, namely, $\kappa$-Minkowski~\cite{lukierski,Dimitrijevic:2003pn,Meljanac:2007xb,Mathieu:2020ywc,Abla:2024wtr,Kurkov:2025soi,Kurkov:2025abv}:
  {\be
[x^{0},x^j]_{\star} = \ii \,\ell \,x^{j}, \qquad [x^j,x^k]_{\star}= 0, \qquad j,k =1,2,3,
\ee}
 the angular noncommutativity, also known as $\lambda$-Minkowski or $\rho$-Minkowski~\cite{Lukierski:2005fc,DimitrijevicCiric:2018blz,Novikov:2019kit,DimitrijevicCiric:2019hqq,Lizzi:2021dud}:
   {\be
[x^3,x^2]_{\star}=\ii\, \ell\, x^1,\quad [x^3,x^1]_{\star}= -\ii\, \ell\, x^2, \quad [x^1,x^2]_{\star}=0,\quad [x^{j},x^0]_{\star}=0,\quad j=1,2,3,
 \ee}
  and the su(2) case~\cite{Hammou:2001cc,Gracia-Bondia:2001ynb,Guedes:2013vi,Vitale:2012dz}:
   { \be
  [x^{i},x^{j}]_{\star}= \ii \,\ell \,\varepsilon^{ijk}\, x^k,\quad [x^{j},x^0]_{\star}=0, \qquad i,j,k =1,2,3,
  \ee}
  are of the Lie-algebra-type:
    {\be
  [x^{\mu},x^{\nu}]_{\star} = \ii\, \mathcal{C}^{\mu\nu}_{\lambda}x^{\lambda}  {.} 
  \ee}
   {The} corresponding star-commutators between the coordinates yield~\eqref{xxcomm} in the semiclassical limit.   {In these formulae, $\ell$ is the fundamental length-scale defining the noncommutativity.}

The presence of the Poisson bracket breaks Poincaré invariance; thus, one may get the impression that these models have essentially fewer symmetries than usual electrodynamics. Our aim is to disprove this naive conclusion. In the present paper, we consider LPE within the geometric approach proposed in~\cite{Kupriyanov:2023qot}. For simplicity, we focus on gauge theory in the absence of charged matter. For \emph{any} continuous symmetry of the Maxwell action, we construct the corresponding deformed symmetry of the LPE action. In particular, we find the deformed Poincaré transformations and the corresponding Noether currents, thereby resolving the open problem of space-time symmetries in LPE, at least for its geometric realization.

  Our construction is essentially based on the field redefinition which maps the LPE and the usual Maxwell dynamics onto each other. Of course, this dynamical equivalence does not imply the physical equivalence, because, as we show in this paper, the local gauge-invariant observables in the two theories are different.

The paper is organized as follows. In Sec.~\ref{SecRev}, we fix the notation and briefly review the relevant aspects of LPE.  In Sec.~\ref{NEWfields}, we construct the field redefinition relating LPE to Maxwell theory and establish the connection between the LPE and Maxwell field equations.   {We also comment on the physical inequivalence of these two field-theoretical models.} In Sec.~\ref{SecSym}, by using the LPE $\leftrightarrow$ Maxwell correspondence, we construct LPE symmetries and Noether currents from their Maxwell counterparts. In Sec.~\ref{SecPoincare}, we apply our findings to Poincaré symmetry. The last section contains a summary of our results, along with interesting quantization perspectives\footnote{  {We emphasize that the proposed quantization prescription should be considered as a  {suggestion} for future research. Our expectations of reasonable properties such as perturbative renormalizability, the lack of a gauge anomaly and unitarity are reasonable    conjectures rather than established results. For a class of LPE with purely spatial noncommutativity we provide additional motivations for our proposal.}  }  naturally arising from the field redefinition proposed in this paper.
Technical details are presented in Appendix~{\bf A}.

\section{Geometric approach to LPE: an overview \la{SecRev}}
In this section, we fix the notation and describe the relevant elements of LPE in general and of the special geometric approach proposed in~\cite{Kupriyanov:2023qot}, mostly following that article and references therein. We focus on explicit expressions in local coordinates and avoid any mention of symplectic groupoids, which are not used in the present paper. The reader interested in the underlying geometric structures is referred to the original reference.  
\subsection*{a. Notations and conventions }
Throughout this article $x^{\mu}$, $\mu =0,.., d-1$, denote the local coordinates on the space-time $\mathcal{M}\simeq \mathbb{R}^d$, and the one-form 
\be
A = A_{\mu}(x)\,\dd x^{\mu}
\ee
is a gauge potential. We assume that $\mathcal{M}$ is equipped with a Poisson bracket 
\be
\{f,g\} = x^{\xi} \,\mathcal{C}^{\mu\nu}_{\xi} \,\partial_{\mu[x]}f\,\partial_{\nu[x]}g, \qquad f,g\in\mathcal{C}^{\infty}(\mathcal{M}),
\ee
where the parameters $\mathcal{C}^{\mu\nu}_{\xi}$ denote structure constants of a given $d$-dimensional Lie algebra $\mathfrak{g}$. The corresponding Lie group, which is unique up to a covering, will be called $G$. Denoting by $p_{\mu}$ the local coordinates on $G$ near its identity element, we introduce bases of left-invariant vector fields and right-invariant one-forms
\be
\gamma^{\mu} =\gamma^{\mu}_{\nu}(p)\,\frac{\partial}{\partial p_{\nu}}, % \la{gammabasis}
\qquad
\rho_{\mu} = \rho_{\mu}^{\nu}(p) \, \dd p_{\nu}. \la{rhobasis}
\ee
The coordinates $p_{\mu}$ are chosen in such a way that 
\be
\lim_{\mathcal{C}\to0} \gamma_{\mu}^{\nu}(p) = \delta_{\mu}^{\nu}  =  \lim_{\mathcal{C}\to0}\rho_{\mu}^{\nu}(p). \la{comlim}
\ee
For a generic Lie algebra $\mathfrak{g}$, suitable expressions for $\gamma^{\mu}_{\nu}$ and $\rho^{\mu}_{\nu}$ have been constructed in~\cite{Kupriyanov:2021cws} and in~\cite{Kupriyanov:2022ohu} respectively in terms of the functions
of the matrix\footnote{According to our convention, upper indices enumerate matrix rows and lower indices enumerate matrix columns.} variable $\hat{p}^{\mu}_{\nu}= \mathcal{C}^{\sigma\mu}_{\nu} p_{\sigma}$,
\be
\gamma(p) = \mathbb{1} +\sum_{k=1}^{\infty}\frac{B^{+}_k\hat{p}^k}{k!}, \, \qquad \rho(p) =  \mathbb{1} +\sum_{k=1}^{\infty}\frac{\hat{p}^k}{(k+1)!} \la{gammarhoexpli}
\ee 
 with $B^{+}_k$,
$k= 1,2,...$, being the Bernoulli numbers.

For any  $Q^{\mu}_{\nu}$, the bar denotes the matrix inverse, $\bar{Q}^{\mu}_{\alpha}\,Q^{\alpha}_{\nu} = \delta^{\mu}_{\nu}$, in particular:
\be
\bar{\rho}^{\nu} = \bar\rho^{\nu}_{\mu}(p)\,\frac{\partial}{\partial p_{\mu}}, \qquad \bar\gamma_{\nu} =\bar\gamma^{\mu}_{\nu}(p)\,\dd{p}_{\mu},  \la{bgammabasis}
\ee
are bases of right-invariant vector fields and left-invariant one-forms on $G$, dual to the bases $\rho_{\mu}$ and $\gamma^{\mu}$ respectively. For the expressions~\eqref{gammarhoexpli} the inverse matrices are simply given by $\bar{\gamma}(p)=\rho(-p)$ and $\bar{\rho}(p)=\gamma(-p)$.

Throughout the paper, we use a single coordinate chart on $G$ defined in a neighborhood of the identity element.
In what follows, the matrices $\gamma^{\mu}_{\nu}(A(x))$, $\rho^{\mu}_{\nu}(A(x))$, and their inverses are understood in this chart, and we restrict ourselves to sufficiently small gauge-field configurations for which $p_{\mu}=A_{\mu}(x)$ remains in this coordinate neighborhood. \\

{\footnotesize \noindent {\bf Remark.} In the physically most interesting case of $d=4$, the above restriction to a single coordinate chart   {can be avoided for almost all Lie algebras $\mathfrak{g}$.} 
Indeed, by choosing $G$ to be the corresponding universal covering group, one finds that, except for $\mathfrak g \cong \mathfrak{su}(2)\oplus\mathfrak{u}(1)$, $G$ is diffeomorphic to $\mathbb{R}^4$, and thereby can be covered globally by a single coordinate chart.  
In the exceptional case, the universal covering group $G\cong SU(2)\times \mathbb{R}$ is diffeomorphic to
$ S^3\times\mathbb{R}$. Using stereographic coordinates on $S^3$, one obtains a single chart covering $G$ everywhere except on a one-dimensional subset.}

\subsection*{b. LPE: the definition and construction }
From now on,  $\delta_f A_{\mu}$ denotes the infinitesimal gauge transformation of $A_{\mu}(x)$ with the gauge parameter $f$.
By definition,  Lie-Poisson electrodynamics is a deformation of Maxwell theory, where the infinitesimal gauge transformations close the non-Abelian algebra
\be
[\delta_f, \delta_g]A_{\mu}(x) = \delta_{\{f,g\}}A_{\mu}. \la{pga}
\ee
In the \emph{commutative} limit $\mathcal{C}\to 0$ of vanishing structure constants, the Poisson bracket on the right-hand side of this equality disappears, and the algebra~\eqref{pga} reduces to the usual $U(1)$ algebra.

The deformed gauge transformations obeying the closure condition~\eqref{pga} can be constructed as follows:
\be
\delta_f A_{\mu}(x) := \gamma_{\mu}^{\nu}\big(A(x)\big)\,\partial_{\nu[x]}f +\{A_{\mu}(x),f(x)\}.  \la{pgt}
\ee 
In the commutative limit, the Poisson bracket vanishes, and thanks to Eq.~\eqref{comlim}, the relation~\eqref{pgt} recovers the standard $U(1)$ gauge transformations:
\be
\lim_{\mathcal{C}\to 0} \delta_f A_{\mu}(x) = \partial_{\mu[x]}f(x).
\ee
In subsequent sections for the gauge variation of $A$ we will use the representation
\be
\delta_ f A_{\mu}(x) = M_{\mu}^{\omega}\,\partial_{\omega[x]} f(x), \la{pgtBIS}
\ee
with
\be
M^{\lambda}_{\varepsilon}:= \gamma^{\lambda}_{\varepsilon}(A(x)) + x^{\xi} \,\mathcal{C}^{\omega\lambda}_{\xi} \,\partial_{\omega[x]}A_{\varepsilon}(x). \la{Mdef}
\ee

The gauge-covariant field-strength tensor is defined by the expression:
\be
F^{s}_{\mu\nu}(x) := \bar\gamma^{\xi}_{\nu}(A)\,\partial_{\mu[x]}A_{\xi} - \bar\gamma^{\xi}_{\mu}(A)\,\partial_{\nu[x]}A_{\xi}  \, {+} \,
\bar\gamma^{\xi}_{\alpha}(A)\,\bar\gamma^{\omega}_{\beta}(A)\,x^{\sigma}\,\mathcal{C}^{\alpha\beta}_{\sigma}\,\partial_{\mu[x]}A_{\xi}\,\partial_{\nu[x]}A_{\omega}. \la{Fs}
\ee
Under the gauge transformations~\eqref{pgt}, the deformed field-strength~\eqref{Fs} transforms by the Lie derivative
\be
\delta_{f} F^{s}_{\mu\nu}(x) =( \mathcal{L}_{\chi_f} F^{s})_{\mu\nu}(x)
\ee
along the Hamiltonian vector field
\be
\chi_f(x) := \{x^{\xi},f\}\,\frac{\partial}{\partial x^{\xi}}. \la{chif}
\ee
Remarkably, the two-form $F^{s} = (1/2)\,F^{s}_{\mu\nu}\, dx^{\mu}\wedge dx^{\nu}$ is exact. In particular, in local coordinates, it can be rewritten as follows:
\be
F^{s}_{\mu\nu}(x) = \partial_{\nu[x]} \big(z^{\alpha}\partial_{\mu[x]} A_{\alpha}\big) - \partial_{\mu[x]} \big(z^{\alpha}\partial_{\nu[x]} A_{\alpha}\big), \la{Fsbis}
\ee
where, by definition,
\be
z^{\mu} := x^{\xi} \,\bar\gamma_{\xi}^{\mu}\big(A(x)\big). \la{zdef}
\ee

In order to construct the gauge-invariant action, one introduces the field-dependent diffeomorphism of $\mathcal{M}$:
\be
y^{\mu}(x) := \Delta^{\mu}_{\xi}\,\big(A(x)\big)\,x^{\xi}, \la{y(x)} 
\ee
with\footnote{The convenient representation~\eqref{y(x)} in terms of the matrix
$\Delta$, defined by Eq.~\eqref{Deltadef}, was presented in~\cite{DiCosmo:2025mme}. For the
matrices $\gamma$ and $\rho$ given by Eq.~\eqref{gammarhoexpli}, one has $\Delta(p)=\exp(-\hat p)$.}
\be
\Delta^{\mu}_{\xi}(p) := \bar{\rho}^{\mu}_{\nu}(p)\, \bar{\gamma}^{\nu}_{\xi}(p). \la{Deltadef}
\ee
  {
The gauge field $A$, its derivatives $\partial A$, the structure constants $\mathcal{C}$ and the coordinates $x$ enter the Jacobian matrix 
\be
J^{\mu}_{\nu} := \frac{\partial y^{\mu}}{\partial x^{\nu}} \la{Jdef}
\ee
through the dimensionless products of the form
$A\,\mathcal C$ and $x \,\partial A\, \mathcal{C}$. In what follows, we shall assume that these combinations are small enough to ensure non-degeneracy of $J$.}
  {Apart from that, we shall assume that the gauge field $A(x)$, being smooth, decays at infinity, which guarantees the global  invertibility\footnote{  {According to Eq.~\eqref{gammarhoexpli}, in this case both matrices $\gamma(A(x))$ and $\rho(A(x))$ tend to the identity matrix at $|x|\to \infty$.  Therefore, in this limit (cf. Eq.~\eqref{Deltadef}) the matrix $\Delta(A(x))$ tends to the identity matrix as well, and hence Eq.~\eqref{y(x)} implies that~$|y(x)| \to \infty$ at $|x|\to \infty$. This property along with the non-degeneracy of $J$ guarantees the global invertibility of our diffeomorphism, see e.g. Theorem 2.2 of~\cite{Ruzhansky}.}} of $y(x)$. }

Let $F^t(y)$ be the pullback of $F^s(x)$ under the inverse map $x=x(y)$:  
\be
F^{t}_{\mu\nu}(y)  = \bar{J}^{\alpha}_{\mu}\,\bar{J}_{\nu}^{\beta}\, F^s_{\alpha\beta}(x(y)). \la{FtDef}
\ee
The key property of this tensor (at fixed $y$) is its gauge-invariance:
\be
\delta_f F^{t}_{\mu\nu}(y) = 0.
\ee 
Therefore, the LPE action,
\be
S[A] = \int_{\mathcal{M}} \dd y  \,\eta^{\mu\alpha}\, \eta^{\nu\beta} \,\Big( -\frac{1}{4}\,F^t_{\mu\nu}(y)F^t_{\alpha\beta}(y) \Big),\la{ginvaction0}
\ee 
with
\be
\eta := \mathrm{diag}\,\big(+1,-1, \cdots,-1\big)
\ee
is a gauge-invariant functional of $A$. Since $F^t(y)$ cannot be expressed locally in terms of $A(y)$,  the corresponding Lagrangian density is not convenient for calculations. Therefore, one performs the change of variables $y\to x$ in the action~\eqref{ginvaction0}:
\be
S[A] = \int_{\mathcal{M}} \dd x \sqrt{ {|g_A|}}\, {g}_A^{\mu\alpha}(x)\,g_A^{\nu\beta}(x)\Big( -\frac{1}{4}\,F^s_{\mu\nu}(x)F^s_{\alpha\beta}(x) \Big)\la{ginvaction1},
\ee
where the field-dependent metric tensor 
\be
g^{A}_{\mu\nu}(x) = J^{\alpha}_{\mu}\,J^{\beta}_{\nu}\eta_{\alpha\beta} \la{gAdef}
\ee
is the pullback of the Minkowski metric $\eta$ under the diffeomorphism~\eqref{y(x)}.
The expression~\eqref{ginvaction1} is gauge-invariant by construction and manifestly local in $A(x)$.

  {Notice that we define the physical space-time intervals by the constant metric $\eta$:
\be
ds^2  = \eta_{\mu\nu} \,dx^{\mu}dx^{\nu}, \la{intervalsdef}
\ee
while the field-dependent metric $g_A$ has to be understood as a technical element of the construction of the gauge-invariant action (2.24). In other words, $x^\mu$ obeying~\eqref{xxcomm} but not $y^{\mu}(x)$ are, \emph{by definition}, {Cartesian} coordinates on a physical flat Minkowski space-time. This  interpretation is important for the discussion of physical     nonequivalence of LPE and Maxwell theory presented in Sec. 3c. 
}

  {In conclusion, we comment  on the regime of dynamical variables which corresponds to the undeformed dynamics.
As we noticed above, the deformation parameters $\mathcal{C}$ enter the Jacobian matrix $J$ through dimensionless contributions of the form $ A\, \mathcal{C}$ and $x\,\partial A\, \mathcal{C}$. Therefore, when  all these combinations are negligible:
\be
| A\, \mathcal{C}| \ll1 , \qquad |x\,\partial A\, \mathcal{C} | \ll 1,  \la{smallcond}
\ee
at leading order  LPE reduces to Maxwell theory. Indeed, in this case, according to Eq.~\eqref{gAdef}:
\begin{equation}
g_{\mu\nu}^A = \eta_{\mu\nu}  + \mathcal{O}(\mathcal{C} A) + \mathcal{O}(x\,\partial A\, \mathcal{C}),
\end{equation}
and  {by} Eq.~\eqref{Fs}: 
\be
F^s_{\mu\nu} = \partial_{\mu}A_{\nu} - \partial_{\nu}A_{\mu} + \cdots,
\ee
where the dots denote corrections of the form $\partial A\, \mathcal{O}(\mathcal{C} A) + \partial A \,  \mathcal{O}(x\,\partial A\, \mathcal{C})$.
Therefore, when the conditions \eqref{smallcond} are satisfied, at the leading order, the action (2.24) reduces to the usual Maxwell action.}

\section{Field redefinition and field equations \la{NEWfields}}
\subsection*{a. Field redefinition}
We start with a simple observation.
The local components of $F^t$ can be represented in the form of the Maxwell field strength:  
\be
F^{t}_{\mu\nu}(y) = F_{\mu\nu}(y),\qquad F_{\mu\nu}(y):=\partial_{\mu[y]}B_{\nu}(y) - \partial_{\nu[y]}B_{\mu}(y), \la{FtF}
\ee
where the one-form $B$ is defined by the relation:
\be
B_{\mu}[A]\big(y(x)\big) := K_{\mu}^{\varepsilon}(x;A)\,A_{\varepsilon}(x), \la{Bdef}
\ee
cf. Eq.~\eqref{Fsbis}, 
and
\be
K_{\mu}^{\varepsilon}(x;A):= \bar{J}^{\beta}_{\mu} \,\, \partial_{\beta[x]} {z}^{\varepsilon}.
\ee
Hereafter, whenever it does not cause confusion, the dependence of the redefined field $B$ on $A$ will be omitted for brevity, and we shall simply write $B_\mu(y)$.

Our main idea is the following. The LPE action introduced in the previous section coincides with the usual Maxwell action for  $B$:  
\be
S[A] = S_{\mathrm{M}}[B],\qquad S_{\mathrm{M}}[B] := \int_{\mathcal{M}} \dd y  \,\eta^{\mu\alpha}\, \eta^{\nu\beta} \,\Big( -\frac{1}{4}\,F_{\mu\nu}(y)F_{\alpha\beta}(y) \Big). \la{corr}
\ee
The continuous symmetries of the Maxwell theory and the associated conserved Noether currents are very well known. We propose to construct the corresponding objects for the LPE by means of our field redefinition~\eqref{Bdef}.

In the commutative limit, the original field $A_{\mu}(x)$ coincides with the new one $B_{\mu}(x)$. 
Hence, at least perturbatively in the structure constants $\mathcal{C}$, one can construct the inverse field redefinition:
\be
A_{\mu}(x) = 
A_{\mu}^{[N]}(x;B)
+\mathcal{O}\big(\mathcal{C}^{N+1}\big), \la{perturbe}
\ee
where 
\be
A_{\mu}^{0}(x;B) := B_{\mu}(x),\qquad  A^{[n]}_{\mu}(x;B):=T^B_{\mu}\big[A^{[n-1]}(\cdot; B)\big](x), \quad n=1,...,N. \la{perturbe1}
\ee
By definition, $T^B$ acts on a smooth one-form $h = h_{\mu}(x)\dd x^{\mu}$ on $\mathcal{M}$ as follows:
\be
T^B_{\varepsilon}[h](x) := \bar{K}_{\varepsilon}^{\mu}(x;h)B_{\mu}(y_h(x)),
\ee
with
\be
y_h^{\mu}(x) := \bar{\rho}^{\mu}_{\nu}\,\big(h(x)\big)\, \bar{\gamma}^{ {\nu}}_{ {\xi}}\big(h(x)\big) \,x^{\xi}.
\ee
The details of this construction are explained in Appendix {\bf A}.
An analysis of the conditions of convergence of the iterative procedure~\eqref{perturbe1} goes beyond the scope of the present paper.

Before elaborating on the deformed symmetries, we establish the connection between the LPE field equations
\be
\mathcal{E}_A^{\mu}(x) = 0, \qquad  \mathcal{E}_A^{\mu}(x) :=\frac{\delta S[A]}{\delta {A_{\mu}(x)}},
\ee
 and their Maxwell counterparts
 \be
 \mathcal{E}_B^{\mu}(y) = 0, \qquad  \mathcal{E}_B^{\mu}(y) :=\frac{\delta S_{\mathrm{M}}[B]}{\delta {B_{\mu}(y)}}.
 \ee

\subsection*{b. LPE and Maxwell field equations}
In order to find this connection, we express the variation $\delta B(y)$ of the field $B$ at \emph{fixed} $y$ through the corresponding variation $\delta A(x)$ 
at fixed $x$. 

On the left-hand side of Eq.~\eqref{Bdef} the argument $y(x)$ also depends on $A$, therefore
\be
\delta \left[B_{\mu}(y(x))\right] = \delta B_{\mu}(y(x)) + \big(\partial_{\nu[y]}B_{\mu}(y)\big)\big|_{y = y(x)} \delta y^{\nu}(x), \la{deltaB}
\ee
where the square brackets mean that $\delta$ acts on the whole expression, including the dependence of $y(x)$ on $A$.
Varying both sides of Eq.~\eqref{Bdef} with respect to $A$, we thus obtain:
 \be
\delta B_{\mu}(y(x)) = - \big(\partial_{\nu[y]}B_{\mu}(y)\big)\big|_{y = y(x)} \delta y^{\nu}(x)+ \delta\big(K_{\mu}^{\varepsilon}(x;A)\,A_{\varepsilon}(x)\big) \la{deltaB1}
\ee
where the ``transport" term $ (\partial_{\nu[y]}B_{\mu})\, \delta y^{\nu}$ has been moved to the right-hand side.

Let us introduce the vector field 
\be
\chi =\chi^{\sigma}(x) \frac{\partial}{\partial x^{\sigma} }, \qquad \chi^{\sigma} := \bar{J}^{\sigma}_\mu\,\delta y^{\mu}. \la{chidef}
\ee 
In Appendix {\bf A} we prove that 
\be
\chi^{\sigma}(x) = x^{\xi}\mathcal{C}^{\sigma\lambda}_{\xi}\,\bar{M}^{\varepsilon}_{\lambda}\,\delta A_{\varepsilon}, \la{chiBIS}
\ee
with  $\bar M$ being the inverse of the matrix $M$ defined by Eq.~\eqref{Mdef}.
In Appendix {\bf A}, we also demonstrate that 
\be
\det{M}= \det{J}\,\big(\det{{\gamma(A)}}\big)^2\,\det{{\rho(A)}}. \la{detM}
\ee
Since $\gamma(A)$ and $\rho(A)$ are invertible by construction, the non-degeneracy of $J$ implies the invertibility of $M$.
According to Eq.~\eqref{chiBIS} and Eq.~\eqref{pgtBIS},  for the gauge variations $\delta A = \delta A_f$, the vector field $\chi$ reduces to the Hamiltonian vector field $\chi_f$, defined by~\eqref{chif}.

After a straightforward calculation,  Eq.~\eqref{deltaB1} can be rewritten as follows:
\be
\delta B_{\mu}(y(x)) = \bar{J}^{\alpha}_{\mu}\,\big(\tilde{\delta}A_{\varepsilon}(x)\,\partial_{\alpha[x]} z^{\varepsilon} 
+ A_{\varepsilon}(x)\,\partial_{\alpha[x]}\tilde{\delta} z^{\varepsilon}\big), \la{deltaB2}
\ee
where we defined the shifted variations 
\be
\tilde{\delta}A_{\varepsilon}(x) :=  {\delta}A_{\varepsilon}(x) - \chi^{\sigma}(x) \,\partial_{\sigma[x]}A_{\varepsilon}(x),\qquad
\tilde{\delta}z^{\varepsilon}(x) :=  {\delta}z^{\varepsilon}(x) - \chi^{\sigma}(x) \,\partial_{\sigma[x]}z^{\varepsilon}(x). \la{tddef}
\ee
 In Appendix {\bf A} we show that
 \be
 \tilde{\delta}A_{\varepsilon}(x) = \gamma^{\lambda}_{\varepsilon}(A(x))\,\bar{M}^{\xi}_\lambda \,\delta A_{\xi}(x),  \la{tdA}
 \ee
 and
\be
\tilde{\delta} z^{\varepsilon}(x) = -z^{\alpha} \,\frac{\partial\gamma_{\alpha}^{\lambda}(A)}{\partial A_{\varepsilon}}\, \bar{M}^{\xi}_\lambda \,\delta A_{\xi}(x).
\la{tdz}
\ee

By substituting the formulae~\eqref{tdA} and~\eqref{tdz} into Eq.~\eqref{deltaB2}, we arrive at the final relation between the variations:
\be
\delta B_{\mu}(y(x))
 = \bar{J}^{\beta}_{\mu}\, \bar{M}^{\sigma}_{\beta}\, \delta A_{\sigma}(x) 
 - \frac{\partial}{\partial y^{\mu}(x)}\bigg(z^{\xi}(x)\,A_{\varepsilon}(x)\frac{\partial\gamma^{\lambda}_{\xi}(A)}{\partial A_{\varepsilon}}\,\bar{M}^{\sigma}_{\lambda}\delta A_{\sigma}(x)\bigg).\la{varirel} 
\ee
  {Hereafter,
\be
\frac{\partial}{\partial y^{\mu}(x)} \equiv \bar{J}^{\beta}_{\mu} \,\frac{\partial}{\partial x^{\beta}}.
\ee
} 

The first variation of the right-hand side of Eq.~\eqref{corr} reads:
\bea
\delta S_{\mathrm{M}}[B] &=& \int_{\mathcal{M}}\dd y\, \mathcal{E}^{\mu}_B(y)\,\delta B_{\mu}(y)   
= \int_{\mathcal{M}}\dd x \,\det{J}\, \mathcal{E}^{\mu}_B(y(x))\,\delta B_{\mu}(y(x)) \nonumber\\
&=& \int_{\mathcal{M}}\dd x \,\det{J}\, \mathcal{E}^{\mu}_B(y(x))\,\bar{J}^{\beta}_{\mu}\, \bar{M}^{\sigma}_{\beta}\, \delta A_{\sigma}(x),   \la{rhsvar}
\eea 
where we performed the change of integration variables and took into account that the total derivative contribution to~\eqref{varirel}, being a gauge transformation of $B$, does not affect the variation of the action. Indeed, undoing the change of variables in the derivative-term, it can be rewritten as follows:
\bea 
\int_{\mathcal{M}}\dd y\,\mathcal{E}^{\mu}_B(y)\, \partial_{\mu[y]}\big(\cdots\big)
 &=& -\int_{\mathcal{M}}\dd y\,\partial_{\mu[y]}\mathcal{E}^{\mu}_B(y)\,\big(\cdots\big) = 0,
\eea
where we integrated by parts and then used the gauge identity for Maxwell theory:
\be
\partial_{\mu[y]}\mathcal{E}^{\mu}_B(y)  = \partial_{\mu[y]}\partial_{\nu[y]} F^{\nu\mu}(y)=0.
\ee

 By construction, the variation~\eqref{rhsvar} must coincide with the variation of the left-hand side of Eq.~\eqref{corr}:
\be
\delta S[A] = \int_{\mathcal{M}}\dd x\, \mathcal{E}^{\sigma}_A(x)\,\delta A_{\sigma}(x),
\ee
for any $\delta A_{\sigma}(x)$, therefore
\be
\mathcal{E}^{\sigma}_A(x) = \det{J}\, \mathcal{E}^{\mu}_B(y(x))\,\bar{J}^{\beta}_{\mu}\, \bar{M}^{\sigma}_{\beta}
\quad \Longleftrightarrow\quad \mathcal{E}^{\mu}_B(y(x)) = (\det{J})^{-1}\,J^{\mu}_{\beta}\, M^{\beta}_{\sigma} \,\mathcal{E}^{\sigma}_A(x)
. \la{dyneq2}
\ee
These relations establish the dynamical equivalence between Maxwell theory and LPE under the non-degeneracy assumptions. In particular, starting from the well-known solutions of the Maxwell equations, one can construct their deformed counterparts perturbatively by using formula~\eqref{perturbe} up to any given order in the deformation parameter.

  {Before we focus on the deformed symmetries, we make an important remark in order to avoid an apparent impression of triviality of LPE.

\subsection*{c. On the physical nonequivalence of LPE and Maxwell theory}
The above result does not imply that LPE is just Maxwell theory rewritten in ``bad variables."  The key point is that the dynamical equivalence does not imply that the two theories describe the same physics, which can be illustrated by a simple classical mechanical example. Consider the  Kepler problem and a two-dimensional isotropic harmonic oscillator. The corresponding Hamiltonians read:
\be
H_{\mathrm{Kepler}} =\frac{p_r^2}{2 m} + \frac{p_{\varphi}^2}{2m r^2} - \frac{k}{r},\qquad H_{\mathrm{Oscil}} =\frac{p_1^2 +p_2^2}{2m} 
+ \frac{m \,\omega^2(q_1^2+q_2^2)}{2},
\ee
where 
$r(t)$ and $\varphi(t)$ are polar coordinates on the orbit plane and $q_1(t)$ and $q_2(t)$ are Cartesian coordinates  in the     configuration space of the harmonic oscillator; $p_r(t)$, $p_\varphi(t)$,  $p_1(t)$ and $p_2(t)$ denote the corresponding conjugate  momenta.  
By introducing the action-angle variables $I_1^{\mathrm{K}}(t), I^{\mathrm{K}}_2(t), \theta^{\mathrm{K}}_1(t),\theta^{\mathrm{K}}_2(t)$ and $I_1^{\mathrm{O}}(t), I^{\mathrm{O}}_2(t), \theta^{\mathrm{O}}_1(t),\theta^{\mathrm{O}}_2(t)$ for these two systems\footnote{  {The normalization of the action-angle  variables $J, w$ in the quoted reference differs from that of our variables $I,\theta$ by the factor of $2\pi$: $J = 2\pi I$, $\theta = 2\pi w$.  We consider the Kepler problem in the plane of the orbit only. Therefore, the action corresponding to the polar angle is zero, whilst $J_r  = 2\pi I_1^{\mathrm{K}}$ and $J_{\varphi} = 2\pi I_2^{\mathrm{K}}$.}}~\cite{Goldstein}, one arrives at dynamical equations of a similar structure:
\be
\frac{\dd I_{1,2}^{\mathrm{K}}}{\dd t} =0,\quad \frac{\dd \theta_{1,2}^{\mathrm K}}{\dd t} =\frac{m k^2}{(I_{1}^{\mathrm{K}}+I_{2}^{\mathrm{K}})^3},\qquad
\frac{\dd I_{1,2}^{\mathrm{O}}}{\dd t} =0,\quad \frac{\dd \theta_{1,2}^{\mathrm O}}{\dd t} =\omega.
\ee
By redefining the evolution parameter $t$ in the two cases as follows:  
\be
\tau_{\mathrm{K}} :=\frac{m k^2\,t}{(I_{1}^{\mathrm{K}}+I_{2}^{\mathrm{K}})^3},\qquad \tau_{\mathrm{O}} :=\omega t \la{timeredef}
\ee
we arrive at the identical equations of motion:
\be
\frac{\dd I_{1,2}}{\dd \tau} =0,\qquad \frac{\dd \theta_{1,2}}{\dd \tau} =1.
\ee
In other words, within the regions of the corresponding phase spaces, where the above transformations are well defined and non-degenerate, the Kepler problem is dynamically equivalent to a two-dimensional isotropic harmonic oscillator. However, this fact does not imply the physical equivalence, because the observer in the former case measures $r(t)$ and $\varphi(t)$, whilst in the latter case the physical observables are $q_1(t)$ and $q_2(t)$, and
on shell $r(t)$ does not coincide with $\sqrt{q_1^2(t) + q_2^2(t)}$.  Moreover,  Eq.~\eqref{timeredef} implies that the dynamical variables of the Kepler problem at the moment of time $t$ are related to those of the harmonic oscillator evaluated at a \emph{different} moment of time $\tilde t= \frac{m k^2 t}{\omega (I_1^K +I_2^K)^3}$.
As in this classical mechanical example, in our field-theoretical setup the redefinition of dynamical variables relates the dynamics of LPE and Maxwell  theory. However, as we explain below, the observables in these two field theories are essentially different, and hence they are not physically equivalent.

In order to compare the local gauge-invariant observables in LPE and Maxwell theory, it is convenient to consider a test-particle moving along the trajectory $x(\tau)$. In the latter case $F_{\mu\nu}(x(\tau))$ is measured.  The LPE case is more subtle, because the test particle's trajectory itself is not gauge-invariant~\cite{Basilio:2024bir}:
\be
\delta_f x^{\mu}(\tau) = \{f(x),x^{\mu}\}\big|_{x=x(\tau)}.
\ee
Thus for any local combination $O(x)$ of $A(x)$ and its derivatives, which transforms (at fixed $x$) by the Poisson bracket:
\be
\delta_{f}{O}(x) = \{O(x),f(x)\}, 
\ee
the expression ${O}(x(\tau))$ is gauge-invariant. For instance, the combination 
\be
\hat{F}_{\mu\nu}(x) := F^t_{\mu\nu}(y(x)) = \bar{J}_{\mu}^{\alpha}\bar{J}_{\nu}^{\beta} F_{\alpha\beta}^s(x)
\ee
transforms properly\footnote{  {As we explained after Eq.~\eqref{detM},  for gauge variations $\delta A = \delta_f A$ the vector field $\chi$ entering Eq.~\eqref{chidef} becomes~$\chi_f$. Therefore, Eqs.~\eqref{chif}, \eqref{Jdef} and~\eqref{chidef}   yield  $\delta_f  y(x) = \{ y(x),f(x)\}$ for the gauge variation of $y(x)$ at fixed $x$. Since $F^t(y)$ is gauge-invariant at fixed $y$, upon the gauge variation of $A$ at fixed $x$ the combination $F^t(y(x))$ transforms by the Poisson bracket:
$\delta _f F^t(y(x)) = \{F^t(y(x)),f(x)\}$.}}:
\be
\delta_f \hat{F}_{\mu\nu}(x)  = \{\hat{F}_{\mu\nu}(x),f(x)\}.
\ee
Therefore $\hat{F}_{\mu\nu}(x(\tau))$ is a legitimate local gauge-invariant observable:
\be
\delta_f\hat{F}_{\mu\nu}(x(\tau)) = 0.
\ee

According to~\eqref{FtF}, the LPE and Maxwell observables are related by
\be
F_{\mu\nu}(y(x(\tau))) = \hat{F}_{\mu\nu}(x(\tau)).
\ee
On the one hand, for a generic field configuration $A(x)$, the gauge-invariant observables  $F_{\mu\nu}(x(\tau))$ and $\hat{F}_{\mu\nu}(x(\tau))$ for the same position of a test particle do not coincide. On the other hand, they take the same value at \emph{different} positions of the test particle, related by the field-dependent diffeomorphism $y(x)$. 
 Summarizing, these noncoinciding space-time positions of a test particle, where the field strengths are evaluated, are a crucial difference between the local gauge-invariant observables in LPE and Maxwell theory.  Interestingly, in our classical mechanical example a similar distinction was present: the change of variables related the Kepler problem and the harmonic oscillator at different moments of time. 
  
 It is important that the above consideration strongly depends on the adopted physical  interpretation of the   coordinates through their relation to measurable intervals.  For example, we assumed that $q_1$ and $q_2$ are \emph{Cartesian} rather than generic  curvilinear coordinates on the configuration space of the two-dimensional harmonic oscillator. Correspondingly (see the end of Sec. 2) it is important that $x^\mu$ are defined as Cartesian    coordinates on a flat Minkowski space-time. For instance, an alternative  choice of $y(x)$ as Cartesian  coordinates on the space time would yield a conclusion that LPE is indeed Maxwell theory rewritten in $A$-dependent  curvilinear   coordinates $x(y)$. However, this choice is not compatible with the adopted definition~\eqref{intervalsdef} of space-time intervals.  {And, of course, by \emph{redefining} the space-time intervals as 
 \be
 ds^2 = \eta_{\mu\nu} \,d y^{\mu}(x)\, dy^{\nu}(x) \la{dsalter}
 \ee
 one may consider LPE and the Maxwell theory physically equivalent at the classical level. Nevertheless, this alternative has its price. For a generic field configuration $A$, the variables $y$ do not close the Lie algebra $\mathfrak{g}$ with respect to the Poisson bracket, 
 \be
 \{y^{\mu}(x),y^{\nu}(x)\} \neq\mathcal{C}^{\mu\nu}_{\xi} y^{\xi}(x). 
 \ee
The left-hand side of this relation, generally speaking, depends  explicitly on $A$ and its derivatives. In other words, the relation between the space-time intervals and the Poisson structure becomes gauge-field dependent and less natural.

}

\section{Deformed symmetries \la{SecSym}}
\subsection*{a. Gauge transformations}
Our field redefinition maps gauge orbits to gauge orbits, that is,   {by definition,} we are dealing with a Seiberg-Witten map~\cite{SW}. Indeed, by setting $\delta A(x) =\delta_ f A(x)$ in Eq.~\eqref{varirel},
and using the representation~\eqref{pgtBIS} for the gauge variation,
 we obtain:
\bea
\delta B_{\mu}(y(x)) &=&  \frac{\partial}{\partial y^{\mu}(x)}\bigg( f(x)
 -z^{\xi}(x)\,A_{\varepsilon}(x)\frac{\partial\gamma^{\lambda}_{\xi}(A)}{\partial A_{\varepsilon}}\,\partial_{\lambda[x]} f(x)\bigg). \la{SW0}
 \eea
This expression is nothing but the Abelian gauge variation
\be
\delta_{\tilde{f}} B_{\mu}(y) = \partial_{\mu[y]}\tilde f(y) \la{SW1}
\ee
with the parameter
\be
\tilde f(y) = \bigg(f(x)
 -z^{\xi}(x)\,A_{\varepsilon}(x)\frac{\partial\gamma^{\lambda}_{\xi}(A)}{\partial A_{\varepsilon}}\,\partial_{\lambda[x]} f(x)\bigg)_{x=x(y)}. \la{SW2}
\ee
The above relation can be inverted at least perturbatively by means of the Neumann expansion:
\be
f(x) = \tilde f(y(x))+\sum_{k=1}^N \mathcal{D}^k \tilde f(y(x)) + \mathcal{O}(\mathcal{C}^{N+1}),
\ee
where $\mathcal{D}$ is the first-order differential operator
\be
\mathcal{D} := z^{\xi}(x)\,A_{\varepsilon}(x)\underbrace{\frac{\partial\gamma^{\lambda}_{\xi}(A)}{\partial A_{\varepsilon}}}_{\mathcal{O}(\mathcal{C})}\,\partial_{\lambda[x]} .
\ee
An analysis of the conditions of convergence of this expansion goes beyond the scope of the present paper. \\

\subsection*{b. Noether currents and associated symmetries}
Let $j^{\mu}_{B}(y)$ be a Noether current of Maxwell theory  associated with the symmetry transformation\footnote{That is, the corresponding variation of the Lagrangian density is a total divergence, and hence the variation of the action vanishes, provided the fields decay sufficiently fast at infinity.}
\be
\delta_u^{\mathrm N} B_{\mu}(y) := R^{B}_{\mu}(y) u, \la{defoBtra}
\ee 
where $R^B$ is the corresponding generator, which depends on $B(y)$ and its derivatives at $y$, and $u$ is an infinitesimal constant parameter. Off-shell we have\footnote{Cf. Eq.~(12) in Noether's original article~\cite{Noether:1918zz}.}
\be
\partial_{\mu}j^{\mu}_{B}(y) = R^{B}_{\mu}(y)\,\mathcal{E}_B^{\mu}(y).
\ee

The corresponding LPE current can be constructed as follows:
\be
j^{\mu}_A(x) := (\det{J})\,\bar{J}^{\mu}_{\nu} j^{\nu}_{B}(y(x)). \la{curr}
\ee 
Indeed, by using the Piola identity for the Jacobian matrix
\be
\partial_{\mu[x]} \big((\det{J})\,\bar{J}^{\mu}_{\nu}\big) =0,
\ee
we get
\bea
\partial_{\mu[x]}j^{\mu}_A(x) &=& (\det{J})\,\bar{J}^{\mu}_{\nu} \partial_{\mu[x]}j^{\nu}_{B}(y(x)) \nonumber\\
&=&(\det{J})\,\big(\partial_{\nu[y]} j^{\nu}_{B}(y)\big)_{y=y(x)} =(\det{J})\,R^{B}_{\mu}(y(x))\,\mathcal{E}_B^{\mu}(y(x)).
\eea
By expressing $\mathcal{E}_B(y(x))$ through $\mathcal{E}_A(x)$ according to~\eqref{dyneq2} we obtain
\be
\partial_{\mu[x]}j^{\mu}_A(x) =  R_{\sigma}^{A}(x)\,\mathcal{E}^{\sigma}_A(x) \la{djA}
\ee
with
\be
R_{\sigma}^{A}(x):=R^{B}_{\mu}(y(x))\,J^{\mu}_{\beta}\, M^{\beta}_{\sigma}. \la{genere}
\ee
By the converse of Noether’s first theorem, Eq.~\eqref{djA} implies that $j^{\mu}_A(x)$ is a Noether current, and $R_{\sigma}^{A}(x)$ are the generators of the corresponding symmetry transformations 
\be
\delta_u^{\mathrm N} A_{\sigma}(x)  := R^{A}_{\sigma}(x)\, u = M^{\beta}_{\sigma}\,J^{\mu}_{\beta}\, \delta^{\mathrm N}_u{B}_{\mu}(y(x)), \la{defotra}
\ee
leaving the LPE action~\eqref{ginvaction1} invariant.

\section{ Deformed Poincaré invariance \la{SecPoincare}}
Below, we analyze in detail an important class of Noether symmetries of the Maxwell action - the Poincaré transformations. The corresponding infinitesimal variations in Eqs.~\eqref{defoBtra} and~\eqref{defotra} will be labeled by P instead of N.
\subsection*{a. Deformed Poincaré transformations}
The infinitesimal Poincaré transformations of $B(y)$ are given by
\be
\delta_{\xi}^{\mathrm{P}}B_{\mu}(y):= (\mathcal{L}_{\xi} {B})_{\mu}(y) \la{PcT}
\ee
where the Lie derivative is taken along the Killing vector field
\be
\xi = \xi^{\mu}(y)\frac{\partial}{\partial y^{\mu}}, \qquad \xi^{\mu}(y) := a^{\mu} +\omega^{\mu\lambda}y_{\lambda}
\ee
of the Minkowski metric $\eta$. In this formula $a^{\mu}$ and $\omega^{\mu\lambda} = -\omega^{\lambda\mu}$ denote the infinitesimal translation and Lorentz transformation parameters respectively,
and
\be
y_{\lambda} :=\eta_{\lambda\nu} y^{\nu}.
\ee 

Eq.~\eqref{defotra} yields the following deformed Poincaré transformations:
\be
\delta_{\xi}^{\mathrm{P}}A_{\sigma}(x) = M^{\mu}_{\sigma}\, \big(\mathcal{L}_{\tilde \xi} {\tilde B}\big)_{\mu}(x),  \la{defPoin}
\ee
where 
\be
\tilde{B} = \tilde{B}_{\mu}(x)\,\dd x^{\mu},\qquad \tilde{B}_{\mu}(x) = J_{\mu}^{\nu}\, B_{\nu}(y(x))
\ee
is simply a pullback of $B$ under the diffeomorphism $y(x)$, whilst
\be
\tilde\xi = \tilde\xi^{\mu}(x)\frac{\partial}{\partial x^{\mu}}, \qquad \tilde\xi^{\mu}(x) = \bar{J}^{\mu}_{\nu} \xi^{\nu}(y(x))
\ee
is a pushforward of $\xi$ under the corresponding inverse map $x(y)$.

It is instructive to separate $\eqref{PcT}$ into two parts: the \emph{covariant} Poincaré transformation and the gauge transformation:
\be
\delta_{\xi}^{\mathrm{P}}B_{\mu}(y) = \delta^{\mathrm{cov}}_{\xi}B_{\mu}(y) +\delta_{\xi}^{\mathrm{gauge}}B_{\mu}(y), 
\ee
where
\be
 \delta^{\mathrm{cov}}_{\xi}B_{\mu}(y) := \xi^{\nu}(y)\,F_{\nu\mu}(y), 
\qquad
\delta_{\xi}^{\mathrm{gauge}}B_{\mu}(y) := \partial_{\mu[y]}\big(\xi^{\nu}(y)B_{\nu}(y) \big). \la{gaut}
\ee

Applying the prescription~\eqref{defotra} to $\delta^{\mathrm{cov}}_{\xi}B$, and using the definition~\eqref{FtDef} of $F^t$ 
together with the relation~\eqref{FtF}, we arrive at the elegant expression for the deformed covariant Poincaré transformations of $A(x)$:
\be
\delta^{\mathrm{cov}}_{\xi}A_{\sigma}(x) = \,M^{\beta}_{\sigma}\,\tilde{\xi}^{\alpha}(x)\,F^s_{\alpha\beta}(x). \la{covtLPE}
\ee
The gauge contribution to~\eqref{gaut} gives rise to the  gauge transformation of $A$ with the  parameter $f(x;\xi):=\xi^{\nu}(y(x))\,B_{\nu}(y(x))$. 
Therefore, the full deformed Poincaré transformation~\eqref{defPoin} in terms of the $F^s$-contribution and the gauge transformation
can be rewritten as:
\be
\delta_{\xi}^{\mathrm{P}}A_{\sigma}(x) =  M^{\beta}_{\sigma}\,\tilde{\xi}^{\alpha}(x)\,F^s_{\alpha\beta}(x)+  \delta_{f(x;\xi)}A_{\sigma}(x).
\ee

\subsection*{b. Conserved currents}
Now, we discuss the conserved currents. For \emph{covariant} translations, that is, for the transformations~$ \delta^{\mathrm{cov}}_{\xi}B$ at $\omega^{\mu\nu}=0$,
we have $d$ conserved currents:
\be
j_{B;(\nu)}^{\mu}(y) = F^{\mu\alpha}(y)\,F_{\nu\alpha}(y)-\frac{1}{4}\,\delta^{\mu}_{\nu} \,  F_{\alpha\beta}(y)\,F^{\alpha\beta}(y) =:-T^{\mu}_{B\,\nu}(y),
\ee
where $T_B$ is the energy-momentum tensor of Maxwell theory. In this formula the indices of $F$ are raised with the Minkowskian metric $\eta$:
\be
F^{\mu\alpha}(y) :=\eta^{\mu\sigma}\eta^{\alpha\omega}F_{\sigma\omega}(y).
\ee
Eq.~\eqref{curr} allows us to construct the LPE Noether currents, which correspond to the deformed transformations~$\delta_{\xi}^{\mathrm{cov}}A$. A simple calculation involving Eq.~\eqref{FtF} and Eq.~\eqref{gAdef} yields the corresponding LPE currents:
\be
j_{A;(\nu)}^{\mu}(x) = (\det{J})\big[\bar{J}_{\nu}^{\zeta}\,F_s^{\mu\alpha}(x)\,F^s_{\zeta\alpha}(x)-\frac{1}{4}\,\bar{J}^{\mu}_{\nu} \,F_s^{\alpha\beta}(x)\,F^s_{\alpha\beta}(x)\big] =:-T^{\mu}_{A\,\nu}(x), \la{TA}
\ee
where $T_A$ is the LPE energy-momentum tensor. The indices of $F_s$ are raised with the metric $g_A$:
\be
F_s^{\mu\alpha}(x):=g_A^{\mu\sigma}g_A^{\alpha\omega}F^s_{\sigma\omega}(x).
\ee
Eq.~\eqref{djA} yields the identity:
\be
\partial_{\mu[x]} j_{A;(\nu)}^{\mu}(x) = \bar{J}^{\alpha}_{\nu}\,M^{\beta}_{\sigma}\,F^s_{\alpha\beta} (x)\,\mathcal{E}^{\sigma}_A(x), 
\ee
which manifests the on-shell conservation of $T_A$.

For \emph{covariant} Lorentz transformations, that is, for~$ \delta^{\mathrm{cov}}_{\xi}B$ at $a^{\mu}=0$, we have $d(d-1)/2$ independent conserved currents:
\be
j^{\mu}_{B;(\alpha,\beta)}(y) = y_{\alpha}\,T_{B\,\beta}^{\mu}(y) - y_{\beta}\,T_{B\,\alpha}^{\mu}(y)  =: M_{B\,\alpha\beta}^{\mu}(y),
\ee
which form the angular-momentum current tensor $M_B$, skew-symmetric in the lower indices. By using the rule~\eqref{curr} along with our previous result~\eqref{TA}, we arrive at the corresponding LPE  currents for the transformations $\delta_{\xi}^{\mathrm{cov}}A$:
\be
j^{\mu}_{A;(\alpha,\beta)}(x) =  y_{\alpha}(x)\,T_{A\,\beta}^{\mu}(x) - y_{\beta}(x)\,T_{A\,\alpha}^{\mu}(x)  =: M_{A\,\alpha\beta}^{\mu}(x), \la{MA}
\ee
which form the LPE angular-momentum current tensor $M_A$. 
Then Eq.~\eqref{djA}, implying the on-shell conservation of these currents becomes:
\be
\partial_{\mu[x]}j^{\mu}_{A;(\alpha,\beta)}(x) =M^{\theta}_{\sigma}\,\big(y_{\beta}(x)\, \bar{J}^{\kappa}_{\alpha}\, F^s_{\kappa\theta} - y_{\alpha}(x)\, \bar{J}^{\kappa}_{\beta}\, F^s_{\kappa\theta}(x) \big)\,\mathcal{E}_A^{\sigma}(x).
\ee

In conclusion, we consider an example of conserved energy in an LPE with a ``purely spatial" Poisson bracket, where the only non-vanishing structure constants are spatial ones:
\be
\mathcal{C}^{\mu\nu}_0 = 0,\qquad  \mathcal{C}^{\mu 0 }_{\nu} = 0.
\ee
One can easily see that in this case $y^0= x^0$, and therefore $\bar{J}^{0}_{\mu}=\delta^{0}_{\mu}$. 
 Denoting by $\bf x$ and $\bf y$ the spatial parts of $x$  and $y$ respectively, and noticing that $\det{J} = \det{(\partial {\bf y}/\partial {\bf x}  )}$, for the conserved energy of the field $A$ we get:
\bea
E[A] &:=& -\int_{\mathbb{R}^{d-1}}\dd {\bf x} \, j_{A;(0)}^0(x) 
=-\int_{\mathbb{R}^{d-1}}\dd {\bf x} \, (\det{J})\,\bar{J}^{0}_{\mu}\,j^{\mu}_{B {;(0)}}(y(x)) \nonumber\\
&=&  -\int_{\mathbb{R}^{d-1}}\dd {\bf y} \, j_{B;(0) }^0(y) = E_\mathrm{M}[B], \la{spatialene}
\eea
that is, the usual expression for the energy $E_\mathrm{M}[B]$ of the Maxwell field $B[A]$.

%%%
\section{Summary and perspectives}

The results of this paper are twofold. On the one hand, we have found deformed symmetries of the geometric realization of LPE. For a generic LPE and for any continuous symmetry of Maxwell theory, we established the generator~\eqref{genere} of the deformed transformation, defining a classical symmetry of the LPE action. The corresponding Noether current is given by Eq.~\eqref{curr}. In particular, applying this general construction to Poincaré symmetry, we obtained the deformed Poincaré transformations~\eqref{covtLPE} of LPE and the corresponding conserved currents~\eqref{TA} and~\eqref{MA}. We notice, however, that the algebra generated by the deformed LPE transformations constructed here remains to be established, and we leave this question as an interesting problem for the future.

On the other hand, our method, namely a field redefinition mapping LPE to Maxwell electrodynamics, is interesting on its own grounds. First, according to 
Eq.~\eqref{SW1} and Eq.~\eqref{SW2}, it provides a Seiberg-Witten map between LPE and $U(1)$ gauge theory. This map is valid for any LPE with gauge transformations~\eqref{pgt}, independently of the structure of the classical action. Apart from the geometric LPE considered in this paper, there is also a ``conventional'' LPE, where the deformed field strength $\mathcal F$ transforms by the Poisson bracket rather than by the Lie derivative,  $\delta_{f} \mathcal{F} = \{\mathcal{F},f\}$, see~\cite{Kupriyanov:2020sgx,Kupriyanov:2023gjj,Kupriyanov:2021cws,Kupriyanov:2022ohu,Kurkov:2025abv} for details. Our field redefinition maps the ``conventional'' LPE dynamics to an Abelian, though nonminimal, nonlinear, and generally speaking Poincaré non-invariant, theory.

Second, our field redefinition yields exciting perspectives for the quantization of LPE. Being highly non-linear, LPE is non-renormalizable by power counting at $d=4$.
However, by introducing the measure $\mathcal{D}B[A]$, formally equal to $\mathcal{D}A\,\det\big(\frac{\delta B}{\delta A}\big)$, in the path integral, one can reduce the problem of calculating the Green's functions
\be
\langle T\, \big[A_{\mu_1}(x_1),...,A_{\mu_n}(x_n)\big]\rangle = Z^{-1} \int\, {\mathcal{D}}B[A]\,T\big[A_{\mu_1}(x_1)\cdots A_{\mu_n}(x_n)\big]\,\mathrm{e}^{\ii S[A]},
  \la{Green}
\ee
with
\be
Z := \int\, {\mathcal{D}}B[A]\,\mathrm{e}^{\ii S[A]}  = \int\, {\mathcal{D}}B \,\mathrm{e}^{\ii S_{\mathrm M}[B]},
\ee
to a more standard problem of calculating the Green's functions of the composite operators in the Maxwell theory. Indeed, by using the perturbative expansion~\eqref{perturbe}, we obtain:
\be
\langle \,T\big[A_{\mu_1}(x_1),...,A_{\mu_n}(x_n)\big]\rangle =  Z^{-1}\int\, {\mathcal{D}}B \,T\big[A^{[N]}_{\mu_1}(x_1;B)\cdots A^{[N]}_{\mu_n}(x_n;B)\big]\, \mathrm{e}^{\ii S_{\mathrm M}[B]} +\mathcal{O}(\mathcal{C}^{N+1}). \la{quantidef}
\ee
At any finite order in $\mathcal{C}$ it is sufficient to substitute $A^{[N]}(x;B)$ by its Taylor ansatz:
\be
A^{[N]}(x;B) = \sum_{n=0}^{N} A^{[N]}_{n}(x;B) + \mathcal{O}(\mathcal{C}^{N+1}), \qquad A^{[N]}_{n}(x;B) = \mathcal{O}(\mathcal{C}^n)
\ee 
where each term $A^{[N]}_{n}(x;B)$ is a local polynomial of $B$ and its derivatives of finite order, evaluated at $x$. 
Eq.~\eqref{quantidef} should be understood as the \emph{definition} of the quantization prescription; in particular, no separate regularization of the
formal Jacobian $\det(\delta B/\delta A)$ is needed. 

Interestingly, the LPE gauge transformations correspond to the Abelian transformations of the field $B$, whilst the measure $\mathcal{D}B$ is gauge-invariant in the Maxwell theory. On the other hand, the naive measure $\mathcal{D}A$ is not gauge-invariant; therefore, according to the logic of Fujikawa~\cite{Fujikawa}, the corresponding naive quantum theory would not be gauge-invariant due to an anomaly. In other words, the path integral measure $\mathcal D B[A] $ seems to be a natural one. 

In the operator language, our choice of the measure corresponds to calculating the vacuum averages over the Maxwell vacuum~$|0\rangle_B$ of the redefined field $B$. This choice is consistent with Eq.~\eqref{spatialene}: for LPE with purely spatial Poisson brackets, the energy of the field $A$ coincides with the Maxwell energy of $B[A]$. This suggests that, for the ``purely spatial'' class of LPE, the Fock space
of the Maxwell field $B$ should provide the natural energy basis for the quantum theory of the field $A$. Of course, a careful canonical quantization requires a constrained Hamiltonian analysis of the geometric LPE. Nevertheless, the above observation gives another physical argument supporting the proposed path-integral prescription, at least for this ``purely spatial'' class of LPE.

In this picture, the quanta of the field $B$ play the role of asymptotic states, whereas the $A$-field is represented at the quantum level by a composite operator $A[B]$. We expect the unitarity of the quantized LPE to be inherited from Maxwell theory, at least for LPE with ``purely spatial'' Poisson brackets, where $y^0=x^0$. A complete quantum treatment should verify this point. We emphasize that the above proposal should be understood as a starting point. A full quantization program requires a careful analysis of gauge fixing, regularization, and renormalization of the composite operators $A^{[N]}_n(x;B)$, where possible tadpole contributions are expected from coincident-point contractions.

It is worth noticing that, although the invertibility of our field redefinition has been proven at the perturbative level only, the construction of deformed symmetries and associated currents uses only the direct map. Moreover, since Green's functions in QFT are typically calculated
perturbatively, perturbative invertibility is precisely what is needed for the above quantization prescription.

Finally, we notice that the geometric approach~\cite{Kupriyanov:2023qot} to LPE is formally applicable to any Poisson bracket, not necessarily of the Lie-algebraic type. Therefore, our idea of constructing the deformed symmetries through a field redefinition can be generalised to a generic Poisson electrodynamics. Indeed, the two-form $F^s$, being closed, is also exact on $\mathbb{R}^d$. Since $F^s$ has the correct commutative limit, it can be presented in the form $F^s=\dd W$, where $W$ can be chosen to have $A$ as its commutative limit.  Then, using the change of variables $y(x)=r_\Sigma(x)$, the redefined field $B$ should be introduced  as\footnote{The diffeomorphism $y(x)=r_\Sigma(x)$ is defined within the symplectic groupoid construction in~\cite{Kupriyanov:2023qot}. For Lie-algebra-type Poisson brackets it becomes~\eqref{y(x)}. Therefore, we use the same notation~\eqref{Jdef} for the Jacobian matrix.}
\be
B_{\mu}(y(x)) = \bar{J}_{\mu}^{\nu} \, W_{\nu}(x).
\ee
The deformed conserved currents can be constructed from their Maxwell counterparts by the same rule~\eqref{curr}, and the generators $R^A_{\sigma}$ of the corresponding deformed transformations should be established from Eq.~\eqref{djA}  as the coefficients of $\mathcal{E}_A^{\sigma}$ on the right-hand side.

\begin{appendix}
\section{Technicalities}
\subsection*{a. On the inverse field redefinition~\eqref{perturbe}}
For a given field configuration $B$, Eq.~\eqref{Bdef} can be rewritten as a fixed-point equation for the operator~$T^B$:  
\be
T^B[A] -A = 0.
\ee
 Below we  show that the ansatz $A^{[N]}$
solves this equation up to terms of order $\mathcal{O}(\mathcal C^{N+1})$: 
\be
T^B[A^{[N]}]-A^{[N]} = \mathcal{O}(\mathcal{C}^{N+1}). \la{rele}
\ee
We assume that the components $B_{\mu}$ are smooth functions of the space-time coordinates and, if they depend on the structure constants
$\mathcal C$, this dependence is non-singular at $\mathcal C=0$.

The expression $T^B[h](x)$ depends on $h$ and its first derivatives, evaluated at $x$. Therefore, its first variation with respect to $h$ is given by
\be
\delta T^B_{\alpha}[h](x):= \bigg(\frac{\partial}{\partial \varepsilon}\,T^B_{\alpha}[h+\varepsilon \,\delta h](x)\bigg)\bigg|_{\varepsilon=0} = \mathcal{S}^{\mu}_{\alpha}[h]\,\delta h_{\mu}(x),
\ee
where $\mathcal{S}$ is the first-order differential operator
\be
\mathcal{S}^{\mu}_{\alpha}[h] = \frac{\partial T^B_{\alpha}[h](x)}{\partial h_{\mu}} + \frac{\partial T^B_{\alpha}[h](x)}{\partial (\partial_{\beta}h_{\mu} )}\, \frac{\partial}{\partial x^{\beta}}. \la{Sdop}
\ee
 For any $h_1,h_2$, the difference $T^B[h_1]-T^B[h_2]$ reads:
\bea
T^B_{\alpha}[h_1](x)-T^B_{\alpha}[h_2](x) &=& \int_{0}^1 \dd s\, \frac{\partial}{\partial s}\, T^B_{\alpha}\big[h_2+s\,(h_1-h_2)\big](x)\nonumber\\
&=&\int_{0}^1 \dd s\,\bigg( \frac{\partial}{\partial \varepsilon}\, T^B_{\alpha}\big[h_2+s\,(h_1-h_2) + \varepsilon\,(h_1 -h_2) \big](x)\bigg)\bigg|_{\varepsilon=0}\nonumber\\
&=&\hat{\mathcal{S}}_{\alpha}^{\mu}[h_1;h_2]\,(h_{1\mu}-h_{2\mu})(x),
\eea
where the first-order differential operator $\hat{S}$ is defined as:
\be
\hat{S}_{\alpha}^{\mu}[h_1;h_2] := \int_{0}^{1} \dd s\,\mathcal{S}^{\mu}_{\alpha}[h_2+s\,(h_1-h_2)].
\ee 
Since  $T^B[h](x)$ is a smooth function of the structure constants $\mathcal{C}$, and 
\be
\lim_{\mathcal{C}\to 0}T^B[h] = B  \la{limiB}
\ee
for any $h$, the coefficients of $\mathcal{S}$  (and hence the coefficients of $\hat{\mathcal{S}}$) 
are of order $\mathcal{O}(\mathcal{C})$ at small $\mathcal{C}$.  Therefore, schematically:
\be
T^B[h_1]-T^B[h_2]=\mathcal{O}(\mathcal{C})\,(h_1-h_2). \la{h1h2}
\ee

Now we prove Eq.~\eqref{rele} by induction. At $N=0$, we get the relation
\be
T^B[A^{[0]}]-A^{[0]} = T^B[B]-B=\mathcal{O}(\mathcal{C}), 
\ee
being an obvious consequence of the commutative limit~\eqref{limiB} at $h =B$.

Supposing that $ T^B[A^{[N-1]}]-A^{[N-1]}=O(\mathcal C^N)$, we have:
\bea
T^B[A^{[N]}]-A^{[N]}
&=&T^B[A^{[N]}]-T^B[A^{[N-1]}] 
=\mathcal{O}(\mathcal C)\,\big(A^{[N]}-A^{[N-1]}\big)  \nonumber\\
&=&\mathcal{O}(\mathcal C)\,\big(T^B[A^{[N-1]}]-A^{[N-1]}\big) 
=\mathcal{O}(\mathcal C)\,\mathcal{O}(\mathcal C^N)
=\mathcal{O}(\mathcal C^{N+1}),
\eea
where we used~\eqref{h1h2} at $h_1 =A^{[N]}$ and $h_2 =A^{[N-1]}$.

\subsection*{b. Derivation of Eq.~\eqref{chiBIS}}
Expanding the right-invariant vector fields $\bar{\rho}^{\nu}$ in the basis of the left-invariant fields $\gamma^{\nu}$ (cf. Eq.~\eqref{Deltadef}) we get:
\be
\bar{\rho}^{\nu}  = \Delta^{\nu}_{\beta}(p)\,\gamma^{\beta},
\ee
therefore
\be
[\gamma^{\mu},\bar\rho^{\nu}]=[\gamma^{\mu},\Delta^{\nu}_{\beta}(p)\,\gamma^{\beta}(p)] = \gamma^{\mu}_{\alpha}(p) \,\frac{\partial \Delta_{\beta}^{\nu}(p)}{\partial p_{\alpha}}\, \gamma^{\beta} +\Delta^{\nu}_{\beta}(p)\,[\gamma^{\mu},\gamma^{\beta}]. \la{intermA}
\ee
Since 
\be
 [\gamma^{\mu},\bar\rho^{\nu}]=0,\qquad [\gamma^{\mu},\gamma^{\beta}] = \mathcal{C}^{\mu\beta}_{\alpha} \,\gamma^{\alpha},
\ee
after a suitable renaming of dummy indices we arrive at
\be
0 = \bigg(\gamma_{\alpha}^{\mu}(p)\,\frac{\partial\Delta_{\beta}^{\nu}(p)}{\partial p_{\alpha}} + \mathcal{C}^{\mu\alpha}_{\beta}\, \Delta_{\alpha}^{\nu}(p) \bigg)\, \gamma^{\beta}.
\ee
The linear independence of $\gamma^{\beta}$ yields the useful formula
\be
\gamma_{\alpha}^{\mu}(p)\,\frac{\partial\Delta_{\beta}^{\nu}(p)}{\partial p_{\alpha}} =  \mathcal{C}^{\alpha\mu}_{\beta}\, \Delta_{\alpha}^{\nu}(p). \la{usefulA}
\ee

By introducing the notation
\be
v_{\lambda} :=\bar{M}^{\varepsilon}_{\lambda}\, \delta A_{\varepsilon}(x), \la{vdef}
\ee
and using the definitions~\eqref{y(x)} and~\eqref{Mdef} we obtain:
\bea
\delta y^{\mu}(x) &=& x^{\nu}\,\frac{\partial\Delta^{\mu}_{\nu}(A)}{\partial A_{\alpha}}\,M^{\lambda}_{\alpha}\,v_{\lambda}\nonumber\\
&=&\bigg( x^{\nu}\, \gamma^{\lambda}_{\alpha}(A)\,\frac{\partial\Delta^{\mu}_{\nu}(A)}{\partial A_{\alpha}}
+ x^{\nu}\,x^{\varepsilon}\,\mathcal{C}^{\varphi\lambda}_{\varepsilon}\,\underbrace{\frac{\partial\Delta^{\mu}_{\nu}(A)}{\partial A_{\alpha}}\,\partial_{\varphi[x]}A_{\alpha}(x)}_{\partial_{\varphi[x]}\Delta^{\mu}_{\nu}(A)}
\bigg)\, v_{\lambda}. 
\eea
Applying Eq.~\eqref{usefulA} to the first term in the right-hand side of this equality, after a suitable renaming of dummy indices, we can rewrite it as follows:
\bea
\delta y^{\mu}(x) &=& x^{\varepsilon}\,\mathcal{C}^{\alpha\lambda}_{\varepsilon}\,\big(\underbrace{\Delta^{\mu}_{\alpha}(A)+x^{\nu}\,\partial_{\alpha[x]}\Delta^{\mu}_{\nu}(A)}_{\partial_{\alpha[x]}y^{\mu}(x) }\big) \,v_{\lambda} \nonumber\\
&=& x^{\varepsilon}\, \mathcal{C}^{\alpha\lambda}_{\varepsilon}\,J^{\mu}_{\alpha}\,v_{\lambda}.
\eea
Multiplying this formula by $\bar J$ we arrive at the desired relation~\eqref{chiBIS}. 
\subsection*{c. Derivation of Eq.~\eqref{detM}}
Introducing two square matrices $U$ and $V$,
\be
U^{\lambda\omega} :=x^{\xi}\,\mathcal{C}^{\omega\lambda}_{\xi},\qquad V_{\omega\sigma} :=\bar{\gamma}^{\varepsilon}_{\sigma}(A)\,\partial_{\omega[x]}A_{\varepsilon}(x),
\ee
and using Eq.~\eqref{Mdef}, we get:
\be
M^{\lambda}_{\varepsilon}\,\bar{\gamma}^{\varepsilon}_{\sigma} = \delta_{\sigma}^{\lambda} + U^{\lambda\varepsilon}\,V_{\varepsilon  {\sigma}},
\ee 
implying the following relation between the matrices:
\be
M\bar{\gamma} = \mathbb{1} + UV. 	\la{inB1}
\ee
On the other hand, the definitions~\eqref{y(x)} and~\eqref{Jdef} along with Eq.~\eqref{usefulA} yield:
\be
J^{\varepsilon}_{\sigma} =\Delta^{\varepsilon}_{\sigma} 
+ x^{\xi}\,\bar{\gamma}^{\omega}_{\theta}(A)\,\mathcal{C}^{\beta\theta}_{\xi}\,\Delta^{\varepsilon}_{\beta}\,\partial_{\sigma[x]}A_{\omega}(x),
\ee
and thus
\be
\bar\Delta^{\lambda}_{\varepsilon}\,J^{\varepsilon}_{\sigma} = \delta^{\lambda}_{\sigma} + U^{\theta\lambda}V_{\sigma\theta},
\ee
that is,
\be
\bar\Delta\,J = \mathbb{1} + U^{\mathrm{T}}V^{\mathrm{T}} = \big(\mathbb{1} + VU\big)^{\mathrm{T}}, \la{inB2}
\ee
where T denotes the matrix transposition.

According to the Weinstein-Aronszajn identity,
\be
\det{\big(\mathbb{1} + UV\big)} = \det{\big(\mathbb{1} + VU\big)}.
\ee
Therefore, Eq.~\eqref{inB1} and Eq.~\eqref{inB2} imply the equality:
\be
\det{(M\bar{\gamma})} =\det{\big(\bar\Delta\,J\big)}.
\ee
Recalling that $\Delta = \bar\rho\,\bar\gamma$, we arrive at~\eqref{detM}.

\subsection*{d. Derivation of Eq.~\eqref{tdA}}
Using the parametrization~\eqref{vdef} of $\delta A$ along with the expression~\eqref{chiBIS} for $\chi$, we obtain:
\bea
\tilde{\delta}A_{\varepsilon}(x) &=& {\delta}A_{\varepsilon}(x) - \chi^{\sigma}(x) \,\partial_{\sigma[x]}A_{\varepsilon}(x) \nonumber\\
&=& M_{\varepsilon}^{\lambda}\,v_{\lambda} - x^{\xi}\mathcal{C}^{\theta\lambda}\,v_{\lambda}\,\partial_{\theta[x]}A_{\varepsilon}(x)\nonumber\\
&=& \gamma^{\lambda}_{\varepsilon}(A)\,v_{\lambda}, \la{Ap1c}
\eea
where we exploited the definition~\eqref{Mdef} at the last step. By expressing $v$ through $\delta A$, we get the desired equation~\eqref{tdA}.
\subsection*{e. Derivation of Eq.~\eqref{tdz}}
From Eq.~\eqref{zdef}, the variation of $z$ reads:
\be
\delta z^{\varepsilon} = x^{\xi}\,\frac{\partial \bar{\gamma}^{\varepsilon}_{\xi}(A)}{\partial A_{\alpha}}\,\delta{A}_{\alpha}(x).
\ee
Therefore, again using the parametrization~\eqref{vdef} of $\delta A$ together with Eq.~\eqref{chiBIS}, we get:
\bea
\tilde{\delta}z^{\varepsilon}(x) = {\delta}z^{\varepsilon}(x) - \chi^{\sigma}(x) \,\partial_{\sigma[x]}z^{\varepsilon}(x)%\nonumber\\
= x^{\xi}\,\bigg(\frac{\partial \bar{\gamma}^{\varepsilon}_{\xi}(A)}{\partial A_{\alpha}}\,M^{\lambda}_{\alpha} 
- \mathcal{C}^{\theta\lambda}_{\xi}\,\partial_{\theta[x]}z^{\varepsilon} \bigg) \, v_{\lambda}.
\eea
Substituting  in this formula the definition~\eqref{Mdef} of $M$ along with 
\be
\partial_{\theta[x]} z^{\varepsilon} = \bar{\gamma}^{\varepsilon}_{\theta}(A) + x^{\sigma}\,\frac{\partial\bar{\gamma}_{\sigma}^{\varepsilon}(A)}{\partial A_{\alpha}}\,\partial_{\theta[x]}A_{\alpha}(x),
\ee
we obtain:
\be
\tilde{\delta}z^{\varepsilon}(x) = x^{\xi}\,\bigg(\gamma_{\alpha}^{\lambda}(A)\frac{\partial \bar{\gamma}_{\xi}^{\varepsilon}(A)}{\partial A_{\alpha}}
-\mathcal{C}_{\xi}^{\theta\lambda}\, \bar{\gamma}_{\theta}^{\varepsilon}(A)
\bigg)\, v_{\lambda} . \la{Ad}
\ee
According to the Maurer-Cartan equation for  $\bar{\gamma}$, expressed in local components,
\be
\frac{\partial \bar{\gamma}_{\xi}^{\varepsilon}(p)}{\partial p_{\alpha}} - \frac{\partial \bar{\gamma}_{\xi}^{\alpha}(p)}{\partial p_{\varepsilon}}
+ \mathcal{C}_{\xi}^{\lambda\theta}\, \bar{\gamma}_{\lambda}^{\alpha}(p)\, \bar{\gamma}_{\theta}^{\varepsilon}(p) = 0,
\ee
the relation~\eqref{Ad} can be rewritten as follows:
\bea
\tilde{\delta}z^{\varepsilon}(x) = x^{\xi}\,\gamma_{\alpha}^{\lambda}(A)\,\frac{\partial \bar{\gamma}_{\xi}^{\alpha}(A)}{\partial A_{\varepsilon}}\,v_{\lambda}
%\nonumber\\
= -x^{\xi}\, \bar{\gamma}_{\xi}^{\alpha}(A)\,\frac{\partial  \gamma_{\alpha}^{\lambda}(A)}{\partial A_{\varepsilon}}\,v_{\lambda} 
=-z^{\alpha}\,\frac{\partial  \gamma_{\alpha}^{\lambda}(A)}{\partial A_{\varepsilon}}\,v_{\lambda}.
\eea
By expressing $v$ through $\delta A$ we complete our proof of Eq.~\eqref{tdz}.
\end{appendix}


\begin{thebibliography}{99}


\bibitem{Kupriyanov:2023qot}
V.~G.~Kupriyanov, A.~A.~Sharapov and R.~J.~Szabo,
``Symplectic groupoids and Poisson electrodynamics,''
JHEP \textbf{03}, 039 (2024)
doi:10.1007/JHEP03(2024)039


\bibitem{Kupriyanov:2020sgx}  
V.~G.~Kupriyanov and P.~Vitale,
``A novel approach to non-commutative gauge theory,''
JHEP \textbf{08}, 041 (2020)
doi:10.1007/JHEP08(2020)041

\bibitem{Kupriyanov:2023gjj}
V.~G.~Kupriyanov, M.~A.~Kurkov and P.~Vitale,
``Lie-Poisson gauge theories and {\ensuremath{\kappa}}-Minkowski electrodynamics,''
JHEP \textbf{11}, 200 (2023)
doi:10.1007/JHEP11(2023)200

\bibitem{Hersent:2022gry}
K.~Hersent, P.~Mathieu and J.~C.~Wallet,
``Gauge theories on quantum spaces,''
Phys. Rept. \textbf{1014}, 1-83 (2023)
doi:10.1016/j.physrep.2023.03.002
%%%%%%%%%%%%%%%%%
%%%%%%%%%%%%%%%%%
% kapparefs
%%%%%%%%%%%%%%%%%
%%%%%%%%%%%%%%%%%
\bibitem{lukierski} 
J. Lukierski, A. Nowicki and H. Ruegg, ``New quantum Poincar\'e algebra, and
$\kappa$-deformed field theory'', Phys. Lett. B293, 344 (1992).


\bibitem{Dimitrijevic:2003pn} 
M.~Dimitrijevic, F.~Meyer, L.~Moller and J.~Wess,
``Gauge theories on the kappa Minkowski space-time,''
Eur. Phys. J. C \textbf{36} (2004), 117-126
doi:10.1140/epjc/s2004-01887-0

\bibitem{Meljanac:2007xb}
S.~Meljanac, A.~Samsarov, M.~Stojic and K.~S.~Gupta,
``Kappa-Minkowski space-time and the star product realizations,''
Eur. Phys. J. C \textbf{53} (2008), 295-309
doi:10.1140/epjc/s10052-007-0450-0


\bibitem{Mathieu:2020ywc}
P.~Mathieu and J.~C.~Wallet,
``Single Extra Dimension from $\kappa$-Poincaré and Gauge Invariance,''
JHEP \textbf{03} (2021), 209
doi:10.1007/JHEP03(2021)209

\bibitem{Abla:2024wtr}
O.~Abla and M.~J.~Neves,
``Poisson electrodynamics on {\ensuremath{\kappa}}-Minkowski space-time,''
Phys. Lett. B \textbf{864}, 139385 (2025)
doi:10.1016/j.physletb.2025.139385

\bibitem{Kurkov:2025soi}
M.~A.~Kurkov,
``Light propagation in $\kappa $-Minkowski space-time: gauge ambiguities and invariance,''
Eur. Phys. J. C \textbf{85}, no.10, 1231 (2025)
doi:10.1140/epjc/s10052-025-14970-9

\bibitem{Kurkov:2025abv}
M.~Kurkov,
``Action principle for {\ensuremath{\kappa}}-Minkowski noncommutative U(1) gauge theory from Lie{\textendash}Poisson electrodynamics,''
J. Phys. A \textbf{59}, no.14, 145402 (2026)
doi:10.1088/1751-8121/ae5776




%%%%%%%%%%%%%%%%
%%%%%%%%%%%%%%%%
% angularRefs 
%%%%%%%%%%%%%%%%
\bibitem{Lukierski:2005fc}
J.~Lukierski and M.~Woronowicz,
``New Lie-algebraic and quadratic deformations of Minkowski space from twisted Poincare symmetries,''
Phys. Lett. B \textbf{633} (2006), 116-124
doi:10.1016/j.physletb.2005.11.052

\bibitem{DimitrijevicCiric:2018blz}
M.~Dimitrijevic Ciric, N.~Konjik, M.~A.~Kurkov, F.~Lizzi and P.~Vitale,
``Noncommutative field theory from angular twist,''
Phys. Rev. D \textbf{98} (2018) no.8, 085011
doi:10.1103/PhysRevD.98.085011
%[arXiv:1806.06678 [hep-th]].

%\cite{Novikov:2019kit}
\bibitem{Novikov:2019kit}
O.~O.~Novikov,
``$\mathcal{PT}$-symmetric quantum field theory on the noncommutative spacetime,''
Mod. Phys. Lett. A \textbf{35} (2019) no.05, 2050012
doi:10.1142/S0217732320500121

\bibitem{DimitrijevicCiric:2019hqq}
M.~Dimitrijevi\'c \'Ciri\'c, N.~Konjik and A.~Samsarov,
``Noncommutative scalar field in the nonextremal Reissner-Nordstr\"om background: Quasinormal mode spectrum,''
Phys. Rev. D \textbf{101} (2020) no.11, 116009
doi:10.1103/PhysRevD.101.116009

\bibitem{Lizzi:2021dud}
F.~Lizzi and P.~Vitale,
``Time Discretization From Noncommutativity,''
Phys. Lett. B \textbf{818} (2021), 136372
doi:10.1016/j.physletb.2021.136372

%%%%%%%%%%%%%%%%
%su2refs
%%%%%%%%%%%%%%%%
\bibitem{Hammou:2001cc}
A.~B.~Hammou, M.~Lagraa and M.~M.~Sheikh-Jabbari,
``Coherent state induced star product on R**3(lambda) and the fuzzy sphere,''
Phys. Rev. D \textbf{66} (2002), 025025
doi:10.1103/PhysRevD.66.025025
%[arXiv:hep-th/0110291 [hep-th]].

\bibitem{Gracia-Bondia:2001ynb}
J.~M.~Gracia-Bondia, F.~Lizzi, G.~Marmo and P.~Vitale,
``Infinitely many star products to play with,''
JHEP \textbf{04} (2002), 026
doi:10.1088/1126-6708/2002/04/026
%[arXiv:hep-th/0112092 [hep-th]].

\bibitem{Guedes:2013vi}
C.~Guedes, D.~Oriti and M.~Raasakka,
``Quantization maps, algebra representation and non-commutative Fourier transform for Lie groups,''
J. Math. Phys. \textbf{54} (2013), 083508
doi:10.1063/1.4818638
%[arXiv:1301.7750 [math-ph]].

\bibitem{Vitale:2012dz}
P.~Vitale and J.~C.~Wallet,
``Noncommutative field theories on $R^3_\lambda$: Toward UV/IR mixing freedom,''
JHEP \textbf{04} (2013), 115
doi:10.1007/JHEP04(2013)115
%[arXiv:1212.5131 [hep-th]].


%%%%%%%%%%%%%%%%
%%%%%%%%%%%%%%%%%

\bibitem{Kupriyanov:2021cws}
V.~G.~Kupriyanov and R.~J.~Szabo,
``Symplectic embeddings, homotopy algebras and almost Poisson gauge symmetry,''
J. Phys. A \textbf{55}, no.3, 035201 (2022)
doi:10.1088/1751-8121/ac411c

\bibitem{Kupriyanov:2022ohu}
V.~G.~Kupriyanov, M.~A.~Kurkov and P.~Vitale,
``Poisson gauge models and Seiberg-Witten map,''
JHEP \textbf{11}, 062 (2022)
doi:10.1007/JHEP11(2022)062


\bibitem{DiCosmo:2025mme}
F.~Di Cosmo, V.~G.~Kupriyanov and P.~Vitale,
``The electromagnetic field in Poisson gauge theory: the groupoidal approach,''
[arXiv:2510.04858 [hep-th]].

%%%%%%%%%%%%%%%%%%%%%%%%%%%%%
\bibitem{Ruzhansky} Ruzhansky, M., Sugimoto, M. ``On global inversion of homogeneous maps,'' Bull. Math. Sci. \textbf{5}, 13–18 (2015). doi:10.1007/s13373-014-0059-1

\bibitem{Goldstein} H. Goldstein, C. Poole, and J. Safko, Classical Mechanics, 3rd ed. (Addison-Wesley, 2002) 

\bibitem{Basilio:2024bir}
B.~S.~Basilio, V.~G.~Kupriyanov and M.~A.~Kurkov,
``Charged particle in Lie{\textendash}Poisson electrodynamics,''
Eur. Phys. J. C \textbf{85} (2025) no.2, 175
doi:10.1140/epjc/s10052-025-13897-5
%7 citations counted in INSPIRE as of 15 Oct 2026
%%%%%%%%%%%%%%%%%%%%%%%%%%%%%

\bibitem{SW}
N.~Seiberg and E.~Witten,
``String theory and noncommutative geometry,''
JHEP \textbf{09}, 032 (1999)
doi:10.1088/1126-6708/1999/09/032

\bibitem{Noether:1918zz}
E.~Noether,
``Invariant Variation Problems,''
Gott. Nachr. \textbf{1918}, 235-257 (1918)
doi:10.1080/00411457108231446
 e-Print: physics/0503066 [physics]

\bibitem{Fujikawa} K. Fujikawa and H. Suzuki, Path integrals and quantum anomalies, Oxford University Press,
Oxford U.K. (2004).





\end{thebibliography}
\end{document}